\input harvmac
\input epsf
\input amssym

\baselineskip 13pt



\newcount\figno
\figno=0
\def\fig#1#2#3{
\par\begingroup\parindent=0pt\leftskip=1cm\rightskip=1cm\parindent=0pt
\baselineskip=11pt
\global\advance\figno by 1
\midinsert
\epsfxsize=#3
\centerline{\epsfbox{#2}}
\vskip -21pt
{\bf Fig.\ \the\figno: } #1\par
\endinsert\endgroup\par
}
\def\figlabel#1{\xdef#1{\the\figno}}
\def\encadremath#1{\vbox{\hrule\hbox{\vrule\kern8pt\vbox{\kern8pt
\hbox{$\displaystyle #1$}\kern8pt}
\kern8pt\vrule}\hrule}}

\def\p{\partial}

\def\rt{\rightarrow}

\def\av{\vec{a}}

\def\zb{\overline{z}}

\def\eps{\epsilon}

\def\Ab{\overline{A}}

\def\zb{\overline{z}}

\def\Wc{{\cal W}}

\def\ab{\overline{a}}

\def\Tb{\overline{T}}

\def\wb{\overline{w}}

\def\Tb{\overline{T}}

\def\Fc{{\cal F}}

\def\Fct{\tilde{\cal F}}
\def\ft{\tilde{f}}

\def\II{\Bbb{I}}


\lref\HadaszGK{
  L.~Hadasz, Z.~Jaskolski and M.~Piatek,
  ``Classical geometry from the quantum Liouville theory,''
Nucl.\ Phys.\ B {\bf 724}, 529 (2005).
[hep-th/0504204].
}

\lref\ChenKPA{
  B.~Chen and J.~J.~Zhang,
JHEP {\bf 1311}, 164 (2013).
[arXiv:1309.5453 [hep-th]].
}

\lref\BarrellaWJA{
  T.~Barrella, X.~Dong, S.~A.~Hartnoll and V.~L.~Martin,
JHEP {\bf 1309}, 109 (2013).
[arXiv:1306.4682 [hep-th]].
}

\lref\AsplundCOA{
  C.~T.~Asplund, A.~Bernamonti, F.~Galli and T.~Hartman,
[arXiv:1410.1392 [hep-th]].
}

\lref\CaputaVAA{
  P.~Caputa, M.~Nozaki and T.~Takayanagi,
PTEP {\bf 2014}, no. 9, 093B06 (2014).
[arXiv:1405.5946 [hep-th]].
}

\lref\CaputaETA{
  P.~Caputa, J.~Simon, A.~Stikonas and T.~Takayanagi,
[arXiv:1410.2287 [hep-th]].
}

\lref\HartmanOAA{
  T.~Hartman, C.~A.~Keller and B.~Stoica,
JHEP {\bf 1409}, 118 (2014).
[arXiv:1405.5137 [hep-th]].
}

\lref\JacksonNLA{
  S.~Jackson, L.~McGough and H.~Verlinde,
[arXiv:1412.5205 [hep-th]].
}

\lref\RobertsIFA{
  D.~A.~Roberts and D.~Stanford,
[arXiv:1412.5123 [hep-th]].
}

\lref\HijanoSQA{
E.~Hijano and P.~Kraus,
``A new spin on entanglement entropy,''
[arXiv:1406.1804 [hep-th]].
}

\lref\zamo{ A. Zamolodchikov,  ``Conformal symmetry in two-dimensional space: recursion represenation of conformal block", Theoretical and Mathematical Physics, Vol. 73, Issue 1, pp 1088-1093, 1987. }

\lref\HartmanMIA{
  T.~Hartman,
  ``Entanglement Entropy at Large Central Charge,''
[arXiv:1303.6955 [hep-th]].
}

\lref\FitzpatrickVUA{
A.~L.~Fitzpatrick, J.~Kaplan and M.~T.~Walters,
``Universality of Long-Distance AdS Physics from the CFT Bootstrap,''
JHEP {\bf 1408}, 145 (2014).
[arXiv:1403.6829 [hep-th]].
}

\lref\DiFrancescoNK{
  P.~Di Francesco, P.~Mathieu and D.~Senechal,
  ``Conformal field theory,''
New York, USA: Springer (1997) 890 p.
}

\lref\HarlowNY{
  D.~Harlow, J.~Maltz and E.~Witten,
JHEP {\bf 1112}, 071 (2011).
[arXiv:1108.4417 [hep-th]].
}

\lref\BelavinVU{
  A.~A.~Belavin, A.~M.~Polyakov and A.~B.~Zamolodchikov,
  ``Infinite Conformal Symmetry in Two-Dimensional Quantum Field Theory,''
Nucl.\ Phys.\ B {\bf 241}, 333 (1984)..
}

\lref\RattazziPE{
  R.~Rattazzi, V.~S.~Rychkov, E.~Tonni and A.~Vichi,
  ``Bounding scalar operator dimensions in 4D CFT,''
JHEP {\bf 0812}, 031 (2008).
[arXiv:0807.0004 [hep-th]].
}

\lref\HeemskerkPN{
  I.~Heemskerk, J.~Penedones, J.~Polchinski and J.~Sully,
  ``Holography from Conformal Field Theory,''
JHEP {\bf 0910}, 079 (2009).
[arXiv:0907.0151 [hep-th]].
}

\lref\FitzpatrickYX{
  A.~L.~Fitzpatrick, J.~Kaplan, D.~Poland and D.~Simmons-Duffin,
JHEP {\bf 1312}, 004 (2013).
[arXiv:1212.3616 [hep-th]].
}

\lref\FitzpatrickCG{
  A.~L.~Fitzpatrick and J.~Kaplan,
JHEP {\bf 1302}, 054 (2013).
[arXiv:1208.0337 [hep-th]].
}

\lref\ElShowkAG{
  S.~El-Showk and K.~Papadodimas,
  ``Emergent Spacetime and Holographic CFTs,''
JHEP {\bf 1210}, 106 (2012).
[arXiv:1101.4163 [hep-th]].
}

\lref\HeemskerkTY{
  I.~Heemskerk and J.~Sully,
  ``More Holography from Conformal Field Theory,''
JHEP {\bf 1009}, 099 (2010).
[arXiv:1006.0976 [hep-th]].
}

\lref\DolanUT{
  F.~A.~Dolan and H.~Osborn,
  ``Conformal four point functions and the operator product expansion,''
Nucl.\ Phys.\ B {\bf 599}, 459 (2001).
[hep-th/0011040].
}

\lref\DolanHV{
  F.~A.~Dolan and H.~Osborn,
  ``Conformal partial waves and the operator product expansion,''
Nucl.\ Phys.\ B {\bf 678}, 491 (2004).
[hep-th/0309180].
}

\lref\ashkin{A. Zamolodchikov, ``Two-dimensional Conformal Symmetry and Critical Four-spin
Correlation Functions in the Ashkin-Teller Model", Zh. Eksp. Teor. Fiz. 90 (1986)
1808{1818.}  }
%

\lref\FaulknerYIA{
  T.~Faulkner,
  ``The Entanglement Renyi Entropies of Disjoint Intervals in AdS/CFT,''
[arXiv:1303.7221 [hep-th]].
}

\lref\AchucarroVZ{
  A.~Achucarro and P.~K.~Townsend,
  ``A Chern-Simons Action for Three-Dimensional anti-De Sitter Supergravity Theories,''
Phys.\ Lett.\ B {\bf 180}, 89 (1986)..
}

\lref\Witten{
 Witten, Edward. ``2+ 1 dimensional gravity as an exactly soluble system." Nuclear Physics B 311.1 (1988): 46-78.
}

\lref\MaldacenaKR{
  J.~M.~Maldacena,
  ``Eternal black holes in anti-de Sitter,''
JHEP {\bf 0304}, 021 (2003).
[hep-th/0106112].
}

\lref\deBoerSNA{
  J.~de Boer, A.~Castro, E.~Hijano, J.~I.~Jottar and P.~Kraus,
  ``Higher Spin Entanglement and W$_N$ Conformal Blocks,''
[arXiv:1412.7520 [hep-th]].
}

\lref\deBoerVCA{
  J.~de Boer and J.~I.~Jottar,
JHEP {\bf 1404}, 089 (2014).
[arXiv:1306.4347 [hep-th]].
}

\lref\AmmonHBA{
  M.~Ammon, A.~Castro and N.~Iqbal,
  ``Wilson Lines and Entanglement Entropy in Higher Spin Gravity,''
JHEP {\bf 1310}, 110 (2013).
[arXiv:1306.4338 [hep-th]].
}

\lref\CastroMZA{
  A.~Castro and E.~Llabrés,
  ``Unravelling Holographic Entanglement Entropy in Higher Spin Theories,''
[arXiv:1410.2870 [hep-th]].
}


\Title{\vbox{\baselineskip14pt
}} {\vbox{\centerline {Worldline approach to semi-classical conformal blocks}}}
\centerline{Eliot Hijano, Per Kraus and River Snively\foot{eliothijano@physics.ucla.edu, pkraus@ucla.edu, river@physics.ucla.edu}}
\bigskip
\centerline{\it{Department of Physics and Astronomy}}
\centerline{${}$\it{University of California, Los Angeles, CA 90095, USA}}

\baselineskip14pt

\vskip .3in

\centerline{\bf Abstract}
\vskip.2cm

\noindent

We extend recent results on semi-classical conformal blocks in 2d CFT and their relation to  3D gravity via the AdS/CFT correspondence.  We consider four-point functions with two heavy and two light external operators, along with the exchange of a light operator.   By explicit computation, we establish precise agreement between these CFT objects and a simple picture of particle worldlines joined by cubic vertices  propagating in asymptotically AdS$_3$ geometries (conical defects or BTZ black holes).   We provide a simple argument that explains this agreement.

\Date{January  2015}
\baselineskip13pt

\newsec{Introduction}

Conformal field theories (CFTs) in two dimensions are specified by a central charge, a list of primary operators, and their OPE coefficients, this data being subject to the consistency requirements of modular invariance and crossing symmetry \refs{\BelavinVU,\DiFrancescoNK}.   The CFT data appears directly in the decomposition of correlation functions in terms of conformal blocks, where the coefficient of each conformal block is given in terms of the OPE coefficients among primary operators.  The conformal blocks correspond to the virtual exchange of a given primary operator and all of its Virasoro descendants.   The conformal blocks are completely fixed by conformal invariance, although no closed form expression for them is known, except  in special cases.  At the same time, efficient recursion relations exist allowing one to compute the conformal block to any desired order in the conformally invariant cross ratio \zamo.

While all of this is ancient history, conformal blocks have received renewed attention recently, both due to their central role in the revival of the conformal bootstrap program \RattazziPE, and also as a useful way to think about how  local  physics can emerge in the bulk in examples of AdS/CFT duality \refs{\HeemskerkPN,\HeemskerkTY,\ElShowkAG,\FitzpatrickCG,\FitzpatrickYX,\FitzpatrickVUA}.\foot{Most of this work is in the context of $d>2$ dimensional CFT, where the conformal group is finite dimensional, and explicit formulas for the conformal blocks are known \refs{\DolanUT,\DolanHV}.}   It is the latter perspective which is closest to our considerations here.  Recent work in this general direction includes the  study of time evolution of entanglement entropy in excited states created by local operators \refs{\CaputaVAA,\CaputaETA,\AsplundCOA}; the issue of universality in 2d conformal field theories \refs{\JacksonNLA,\HartmanOAA}; entanglement entropy in higher spin theories \deBoerSNA; and the emergence of chaos in thermal systems \RobertsIFA.

We are interested in conformal blocks in the semi-classical limit, corresponding to taking the central charge and operator dimensions to infinity while keeping their ratios fixed.  There is excellent evidence that in this limit the conformal blocks exponentiate \refs{\ashkin,\HarlowNY}  as $e^{-{{c}\over{6}} f({h_i \over c}; x)}$, although there exists no  proof of this directly from the fundamental definition as a sum over Virasoro descendants.    It is then natural to expect that the function $f({h_i \over c}; x)$ can be computed from a saddle point analysis, and indeed several realizations of this are known, as we discuss.

Since conformal blocks are fixed by conformal symmetry, they can be computed in any theory with this symmetry, provided of course that the desired values of the central charge and operator dimensions are available.   If we have a family of theories with a variable central charge, where $1/c$ plays the role of $\hbar$,  then we can think of taking $c \rt \infty$ and computing $f({h_i \over c}; x)$ by solving some ``classical equations".   An example is provided by Liouville theory, and this motivates the monodromy approach to the computation of semi-classical conformal blocks \refs{\ashkin,\HarlowNY}.   This approach boils down to solving an ordinary differential equation, $\psi''(z) +T(z) \psi(z)=0$, with prescribed monodromy.   The monodromy condition fixes certain ``accessory parameters" in $T(z)$, and these can be used to reconstruct the conformal block by integration.

Another useful realization is in terms of gravity in AdS$_3$.   As was emphasized in the context of holographic entanglement entropy \refs{\HartmanMIA,\FaulknerYIA}, the monodromy approach can be recast in terms of the problem of finding a solution of Einstein's equations with specified boundary behavior.  In this context, $T(z)$ is identified as the boundary stress tensor of AdS$_3$ gravity. The authors of \refs{\HartmanMIA,\FaulknerYIA} were thereby able to derive from first principles the Ryu-Takayanagi formula in this context, equating the entanglement entropy with the (regulated) length of a bulk geodesic.

Especially relevant for our purposes is the illuminating paper \FitzpatrickVUA\ which, among other things, gave an AdS$_3$ bulk interpretation of vacuum conformal blocks for the case in which two of the external operators are heavy and two are light, in a sense made precise below.   The picture is that the heavy operators set up a classical asymptotically AdS$_3$ geometry corresponding to a conical defect or BTZ black hole, and the light operators are described by a geodesic probing this background solution.

Here we extend the results of    \FitzpatrickVUA\ in several directions.   Firstly, we  consider the case of nonvacuum conformal blocks corresponding to an exchanged primary $O_p$, and also allow for the light operators to have distinct conformal dimensions.  The bulk picture is that we now have three geodesic segments in the background geometry, one segment for each of the light operators, with the segments meeting at a cubic vertex; see figure 1.  The segment corresponding to the operator $O_p$ has one endpoint on the vertex, with the other ending at the conical defect or BTZ horizon.    The resulting formula for the conformal block is  more intricate than the case considered in \FitzpatrickVUA, but is still quite compact.  By expanding this result for small cross ratio, we are able to check against the corresponding result obtained directly from the CFT recursion relation, and we indeed find agreement.   We also verify agreement with the monodromy approach.

\vskip-.3cm
\fig{Geodesic configuration.   The disk represents a  slice of a conical defect or BTZ geometry, as supported by the heavy operator.   Light operators are represented by geodesic segments as shown.}{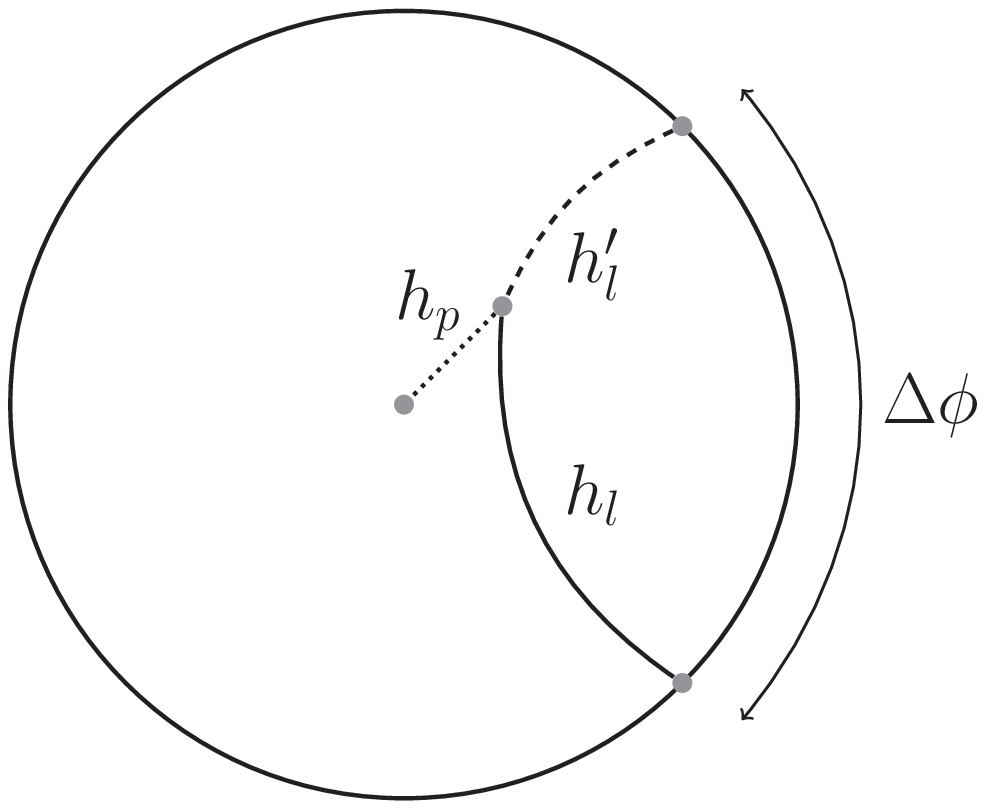}{3.0truein}

We  also  provide a simple argument that explains the agreement between the bulk geodesic approach and the monodromy approach.   This is done by thinking about the backreaction on the metric produced by the configuration of particle geodesics.   The relation to the CFT monodromy approach is especially transparent in the Chern-Simons formulation of AdS$_3$ gravity \refs{\AchucarroVZ,\Witten}, and this lets us establish that the solution sourced by the geodesics is in direct correspondence with a solution of the monodromy problem.

The geodesic approximation is only valid to first order in the light operator dimensions, while the full conformal block of course receives contributions at all orders.  In the bulk we can think about solving Einstein's equations order by order to compute these corrections.   Equivalently, this can  be phrased in terms of solving the monodromy problem at higher orders.  We verify that the second order solution  indeed yields a result in agreement with that obtained from the recursion relation.

Before proceeding, we would like to emphasize that a motivation for carrying out the work presented here is to eventually apply these results to computations that are not entirely dictated by symmetry.  For instance, semi-classical correlation functions computed in the BTZ geometry display a specific form of information loss \MaldacenaKR.  In the present context such correlation functions arise in a manner in which it is clear what effects have been thrown out, namely non-vacuum blocks and $1/c$ corrections, and this might be a useful way to think about what is needed to restore purity. For a calculation of one loop corrections of holographic entanglement entropy, see \BarrellaWJA.

\newsec{Conformal blocks}

In this section we briefly review the definition of conformal blocks in 2D CFT, as well as some of the methods available to compute them, either in a series expansion or in the semi-classical limit.

\subsec{Definitions}

We follow the conventions of \DiFrancescoNK.  The correlation function of four primary operators is expanded as
\eqn\aa{ \langle O_1(\infty,\infty)O_2(1,1) O_3(x,\overline{x})O_4(0,0)\rangle =\sum_p C^p_{34} C^p_{12} \Fc^{21}_{34}(p|x)\overline{\Fc}^{21}_{34}(p|\overline{x})~,}
where $ O_1(\infty,\infty) = \lim_{z_1, \overline{z}_1 \rt \infty}z_1^{2h_1} \overline{z}_1^{2 \overline{h}_1} O_1(z_1,\overline{z}_1)$ inside the correlator.
The expansion \aa\ is obtained by using the $O_1 O_2$ and $O_3 O_4$ OPEs, together with the fact that the OPE coefficients involving Virasoro descendants are related by conformal symmetry to those of the primaries. Each term in \aa\ corresponds to the virtual exchange of a primary $O_p$ together with all of its Virasoro descendants.

The conformal block $\Fc^{21}_{34}(p|x)$ admits a series expansion,
\eqn\ab{ \Fc^{21}_{34}(p|x)=x^{h_p-h_3-h_4}\Fct^{21}_{34}(p|x)~,\quad  \Fct^{21}_{34}(p|x)=\sum_{n=0}^\infty [\Fct^{21}_{34}]_n x^n }
with $[\Fct^{21}_{34}]_0=1$.   At the next two orders we have\foot{Note that 6.191 in \DiFrancescoNK\ is incorrect; see the errata.}
\eqn\ac{\eqalign{\!\!\!\!& [\Fct^{21}_{34}]_1 = {(h_p+h_2-h_1)(h_p+h_3-h_4)\over 2h_p} \cr
&[\Fct^{21}_{34}]_2 =  {A+C\over B} \cr
&A = (h_p+h_2-h_1)(h_p+h_2-h_1+1) \Big[(h_p+h_3-h_4)(h_p+h_3-h_4+1)(4h_p +{c\over 2})\cr
& \quad\quad\quad\quad\quad\quad\quad\quad\quad\quad\quad\quad\quad\quad\quad\quad\quad -6h_p(h_p+2h_3-h_4)\Big]  \cr
&C = (h_p+2h_2-h_1)\Big[4h_p(2h_p+1)(h_p+2h_3-h_4)-6h_p(h_p+h_3-h_4)(h_p+h_3-h_4+1\Big]\cr
&B = 4h_p(2h_p+1)(4h_p+{c\over 2})-36h_p^2   }}
Higher order terms are readily computed using a convenient recursion relation \zamo.

\subsec{Heavy-light correlators, and the semi-classical limit}

It will now be convenient to define a rescaled stress tensor and rescaled conformal weights.  If $T_{CFT}$ and $h$ denote the usual stress tensor and conformal weight, we now define $T$ and $\eps$ as
\eqn\ad{ T_{CFT}(z) = {c\over 6}T(z)~,\quad h= {c\over 6}\eps,}
In terms of which the OPE is
\eqn\ae{ T(z) O(0) = {\eps \over z^2} O(0)+{{6}\over{c}} {1\over z}\p O(0) +\ldots~.}
The semi-classical limit of the conformal blocks  is defined by taking $c \rt \infty$ at fixed $\eps$, where $\eps$ refers to the external operators $O_{1,2,3,4}$ as well as the internal primary $O_p$. In this limit there is good evidence, though no direct proof, that the conformal blocks exponentiate
\eqn\af{ \Fct^{21}_{34}(p|x) = e^{-{c\over 6}\ft^{21}_{34}(\eps_i;x)}~.}
This can be verified directly to the first few orders in the $x$ expansion using the recursion relation obtained in \zamo\ and reviewed in appendix A. It has been verified to high order in \HadaszGK.
We will be interested in the case in which the four-point function involves two operators of equal dimension $\eps_h$, with the other two operators having dimensions $\eps_l$ and $\eps'_l$.  Here the subscripts stand for ``heavy" and ``light", in the sense that we will be consider $\eps_l, \eps'_l \ll 1$ so that we can expand perturbatively in these quantities.  Note though that our usage of ``light" is somewhat nonstandard, since this label is often applied to operators whose dimension $h$ is held fixed as $c\rt \infty$, which is not the case here.

In particular, the correlator of interest is (suppressing henceforth the dependence on anti-holomorphic quantities)
\eqn\ag{  \langle h |  O_{l'}(1) O_l(x)|h\rangle =\langle O_h(\infty) O_{l'}(1) O_l(x) O_h(0)\rangle~,  }
expanded in the $x \rt 1$ OPE channel.   As indicated, we can think of this correlator as  the two-point function of light operators in the excited state created by the heavy operators.  To expand in conformal blocks, first use invariance under $z\rt 1-z$ to write
\eqn\ah{ \langle O_h(\infty) O_{l'}(1) O_l(x) O_h(0)\rangle  = \langle O_h(\infty)O_h(1) O_l(1-x) O_{l'}(0)\rangle}
so that the expansion is
\eqn\ai{ \langle h |  O_{l'}(1) O_l(x)|h\rangle = \sum_p C^p_{ll'}C^p_{hh}\Fc^{hh}_{ll'}(p|1-x)\overline{\Fc}^{hh}_{ll'}(p|1-\overline{x})~.}
Suppressing the labels, we then write, in the semi-classical limit,
\eqn\aj{ \Fc(1-x) = (1-x)^{h_p-h_l-h'_l}\Fct(1-x) = (1-x)^{{c\over 6}(\eps_p-\eps_l-\eps'_l)}e^{-{c\over 6}\ft(1-x)}~.}

As noted above, the recursion relation can be used to compute $\ft(x)$ in a power series, although the results rapidly get complicated.   To simplify we further expand $\ft(x)$ in the light operator dimensions. More precisely, we replace
\eqn\ak{ \eps_l \rt \delta \eps_l~,\quad \eps'_l \rt \delta \eps'_l~,\quad \eps_p \rt \delta \eps_p}
and then expand in $\delta$.   At linear order in $\delta$ we find
\eqn\al{\eqalign{
\ft_\delta(x) &= {1\over 2}(\eps'_l-\eps_l-\eps_p)  x  \cr
&-{1\over 16} {(1-4\eps_h)(\eps'_l-\eps_l)^2 \over\eps_p} x^2 + \left[{\eps'_l-\eps_l\over 4} -{(\eps'_l+\eps_l)\eps_h\over 6}  \right] x^2- \left[{\eps_h\over 12}+{3\over 16} \right] \eps_p x^2  \cr
&-{1\over 16} {(1-4\eps_h)(\eps'_l-\eps_l)^2 \over\eps_p}x^3 + \left[{\eps'_l-\eps_l\over 6} -{(\eps'_l+\eps_l)\eps_h \over 6}  \right]x^3- \left[{\eps_h\over 12}+{5\over 48} \right] \eps_p  x^3\cr
& + \ldots }}
while at quadratic order in $\delta$ we have
\eqn\am{\eqalign{ \ft_{\delta^2}(x)& =
 -{1\over 48}{(16\eps_h-3)(\eps'_l-\eps_l)^2}x^2+{1\over 72}(16\eps_h-3)(\eps'_l+\eps_l)\eps_p x^2 +{1\over 144}(16\eps_h-3)\eps_p^2 x^2 \cr
& -{1\over 48}{(16\eps_h-3)(\eps'_l-\eps_l)^2}x^3+{1\over 72}(16\eps_h-3)(\eps'_l+\eps_l)\eps_p x^3 +{1\over 144}(16\eps_h-3)\eps_p^2 x^3\cr
& + \ldots ~.}}
The expansions continue to higher orders in $x$ and $\delta$.

\subsec{Monodromy method}

A convenient method for computing the semi-classical conformal block is the monodromy method.  This is well reviewed in \refs{\HarlowNY,\FitzpatrickVUA}, and so we will be brief.   We consider the differential equation
\eqn\an{ \psi''(z) +T(z) \psi(z) =0 }
with
\eqn\ao{T(z) =  {\eps_h \over z^2} + {\eps'_l \over (z-1)^2}+ {\eps_l\over (z-x)^2} + {\eps_l+\eps'_l \over z(z-1)}+{x(1-x) \over z(1-z)(z-x)}c_x(x) }
$T(z)$ can be thought of as the stress tensor in the presence of the operators appearing in the four-point function \ag.  Up to the free parameter $c_x$, its form is fixed by demanding that it have the correct double poles and asymptotic behavior.  Noting that the simple pole term at $z=x$ is $T(z) \sim {c_x \over z-x} $, along with the OPE \ae, we see that $c_x$ is related to the $x$-derivative of the conformal block.   We can then integrate as
\eqn\ap{ f(1-x) = -\int\! c_x(x) dx}
where $\Fc= e^{-{c\over 6}f}$ in the semi-classical limit.

$c_x(x)$ is determined by demanding that the two independent solutions of \an\ undergo a specific monodromy as we go around a contour that encloses the singularities at $z=1, x$, the monodromy being fixed by the dimension of the primary $\eps_p$.   Specifically, the monodromy matrix $M$ is required to have eigenvalues
\eqn\aq{ \lambda_\pm = e^{i\pi(1\pm \sqrt{1-4\eps_p})}~.}%

The problem is tractable in perturbation theory, where we can expand in the dimensions of the light operators, or  in $1-x$.   For the former we write
\eqn\ar{\eqalign{ \psi&= \psi^{(0)}+\psi^{(1)} + \psi^{(2)} +\ldots \cr
 T(z) &= T^{(0)} +T^{(1)} +T^{(2)} + \ldots }}
with
\eqn\as{\eqalign{ T^{(0)}& = {\eps_h \over z^2}  \cr
T^{(1)}& =  {\eps'_l \over (z-1)^2}+ {\eps_l\over (z-x)^2} + {\eps_l+\eps'_l \over z(z-1)}+  {x(1-x) \over z(1-z)(z-x)}c^{(1)}_x\cr
T^{(2)}&=  {x(1-x) \over z(1-z)(z-x)}c^{(2)}_x }}
The equations are then
\eqn\at{\eqalign{ & (\psi^{(0)})''+T^{(0)}\psi^{(0)}=0  \cr
& (\psi^{(1)})''+T^{(0)}\psi^{(1)}=-T^{(1)}\psi^{(0)} \cr
 & (\psi^{(2)})'' +T^{(0)}\psi^{(2)}   = -T^{(1)}\psi^{(1)}-T^{(2)}\psi^{(0)}  }}
The zeroth order solutions are
\eqn\au{ \psi^{(0)}_\pm(z) = z^{{1\pm \alpha \over 2}}~,\quad \alpha = \sqrt{1-4\eps_h} }
We then obtain solutions at the next two orders as
\eqn\av{\eqalign{  \psi_\pm^{(1)}(z) &=   \left[-{1\over \alpha}\int^z\! dz {\psi_-^{(0)} T^{(1)} \psi_\pm^{(0)} }\right]\psi_+^{(0)}(z)+  \left[{1\over \alpha} \int^z\! dz {\psi_+^{(0)} T^{(1)} \psi_\pm^{(0)} \over W}\right]\psi_-^{(0)}(z) \cr
 \psi_\pm^{(2)}(z) & = \left[ -{1\over \alpha}  \int^z\! \psi_-^{(0)} \big( T^{(1)}\psi_\pm^{(1)}+T^{(2)}\psi_\pm^{(0)}  \big)\right]\psi_{+}^{(0)}(z) \cr
 & \quad \quad+ \left[{1\over \alpha} \int^z\! \psi_+^{(0)} \big( T^{(1)}\psi_\pm^{(1)}+T^{(2)}\psi_\pm^{(0)}\big)\right]\psi_-^{(0)}(z) }}
In this form it is easy to read off the monodromy matrix as we encircle $z=1, x$.  Writing $M=M^{(0)} + M^{(1)}+M^{(2)} + \ldots$ we have $M^{(0)}= \II$ and
\eqn\aw{\eqalign{M^{(1)}_{++} & = -{1\over \alpha}\oint\! dz {\psi_-^{(0)} T^{(1)} \psi_+^{(0)} }~,\quad   M^{(1)}_{+-}  = {1\over \alpha}\oint\! dz {\psi_+^{(0)} T^{(1)} \psi_+^{(0)} } \cr
M^{(1)}_{-+} & = -{1\over \alpha}\oint\! dz {\psi_-^{(0)} T^{(1)} \psi_-^{(0)} }~,\quad   M^{(1)}_{--}  = {1\over \alpha}\oint\! dz {\psi_+^{(0)} T^{(1)} \psi_-^{(0)} }}}
and similarly for $M^{(2)}$.   The integrals are easily computed using residues.   Now let us give a few examples.

\vskip.2cm
\noindent
$\bullet$ First order in $ \eps'_l=\eps_l, \eps_p $
\noindent
\vskip .2cm

This case was consired in \FitzpatrickVUA.  Using $M^{(1)}_{++} = M^{(1)}_{--}=0$, the equation determining $c_x$ is
\eqn\ax{ M^{(1)}_{+-}M^{(1)}_{-+} =-4\pi^2 \eps_p^2~, }
yielding
 \eqn\ay{c_x = -\left[{1\over x} + {\alpha\over x} \left({x^{{\alpha\over 2}} +x^{-{\alpha\over 2}}  \over  x^{{\alpha\over 2}} -x^{-{\alpha\over 2}}}\right)   \right]\eps_l  + {\alpha \over x (x^{{\alpha\over 2}} -x^{-{\alpha\over 2})}}\eps_p }
This gives (choosing the integration constant so that $\ft(0)=0$)
\eqn\az{ f_\delta(1-x)= \left[\ln x +2 \ln \left( x^{-{\alpha\over 2}}- x^{{\alpha\over 2}}\over \alpha\right)  \right] \eps_l -  \ln \left( {x^{-{\alpha\over 4}}-x^{{\alpha\over 4}}\over 4\alpha( x^{-{\alpha\over 4}}+x^{{\alpha\over 4}})} \right)\eps_p  }
Using
\eqn\ba{ \ft_\delta(x) = f(x) +(\eps_p -\eps_l-\eps'_l)\ln x }
and expanding in $x$ we verify agreement with \al.

\vskip.2cm
\noindent
$\bullet$ First order in $ \eps'_l\neq \eps_l, \eps_p $
\noindent
\vskip .2cm

We now allow for two independent light operators.  The result turns out to agree precisely with (3.30), the latter being obtained from a bulk computation.

\vskip.2cm
\noindent
$\bullet$ Second order in $ \eps'_l=\eps_l$ with $\eps_p =0$
\noindent
\vskip .2cm

In this case we solve $M^{(2)}=0$.   In general this is quite complicated due to the complexity of the second order solution $\psi^{(2)}$.   To give a very simple illustration, we choose the light operators as above, and further expand to lowest nontrivial order in $1-x$, which gives
\eqn\bb{ c^{(2)}_x =  \left[-{2 \over 45}\eps_h+{22  \over 135}\eps_h^2\right] \eps_l^2 (1-x)^3+\ldots}
which yields
\eqn\bc{ \ft_{\delta^2}(x) =f_{\delta^2}(x)  =    \left[-{1\over 90}\eps_h +{11\over 270} \eps_h^2  \right] \eps_l^2  x^4 +\ldots }
This matches the result from the recursion relation (the term one order beyond those given in \am).   It is simple to extend to higher orders in $x$, if desired.

\newsec{Semi-classical conformal blocks from bulk geodesics}

In \FitzpatrickVUA\ it was observed that the conformal block with $\eps'_l=\eps_l$ and $\eps_p=0$ can be reproduced by computing the length of a geodesic in an asymptotically AdS$_3$ background.   In this section we show how to extend this to reproduce the conformal block for general values of $\eps'_l$, $\eps_l$, $\eps_p$, assuming all are small.

\subsec{Setup}

It is conceptually easiest work in global coordinates corresponding to CFT on the cylinder.  The geometry related to the operator $O_h$ by the state-operator correspondence is
\eqn\bd{ ds^2 = {\alpha^2 \over \cos^2 \rho}( {1\over \alpha^2} d\rho^2 -dt^2 +\sin^2 \rho d\phi^2) }
which is obtained from global AdS$_3$ by $t \rt  \alpha t $, $\phi \rt \alpha \phi$, although in \bd\ we take $\phi \cong \phi+2\pi$.   Here $\alpha$ is the same quantity as in \au.  For $\alpha^2>0$ \bd\ represents a conical defect with a singularity at $\rho =0$.  We define $w=\phi+it_E$, with $t_E=it$.

For $\alpha^2  <0$ we instead have a BTZ black hole with event horizon at $\rho=0$, provided that we change the identifications to $t\cong t+2\pi$.  Since we take $\phi\cong \phi+2\pi$, we will mainly restrict to the $\alpha^2>0$ case.

Now consider computing AdS correlation functions of the operators $O_l$ and $O_{l'}$.  The light operators are placed at:
\eqn\bda{  O_{l'}(w=0)~,\quad O_{l}(w)~. }
We work on a fixed time slice of the conical defect, so that $w=\phi$.
 An operator of dimension $(h,h)$ is dual to a bulk field of mass $m= 2\sqrt{h(h-1)}$. Since we are always assuming $h\sim c \gg 1$, this is $m = 2h \gg 1$.  In this regime, the scalar field is well approximated by a point particle, and correlation functions can be computed in terms of regulated geodesic lengths. To compute the conformal block involving the exchanged primary $O_p$ we propose the following simple prescription, which we will subsequently verify in particular cases, and then justify on general grounds.  Working on a fixed $t$ slice, we take the external light operators to be inserted as in \bda.    Attached to the boundary point $w=0$ is a segment of geodesic corresponding to a particle of mass $m_{l'} = 2h_{l'}$. Similarly, a $m_l=2h_l$ geodesic segment is attached at $w$.   The primary operator is represented by a geodesic segment attached at $\rho=0$ in the background \bd.   The three geodesic segments meet at a cubic vertex, located at some point in the interior of \bd.   The worldline action is, after stripping off a factor of ${c\over 6}$,
\eqn\bez{ S = \eps_{l}'L_{l'} + \eps_l L_l + \eps_p L_p }
where the $L$s denote the regulated lengths of the geodesics.\foot{Note that it is $h$ that appears here rather than $m=2h$, since we are computing the chiral half of the correlator.}  The location of the cubic vertex is obtained by minimizing $S$.       This yields a contribution to the correlation function on the cylinder
\eqn\bfz{ G(w) = e^{-{c\over 6} S(w)}~. }
This is related to the conformal block on the plane by the conformal transformation $z=e^{iw}$,
\eqn\bg{ \Fc(1-z) = z^{-h_l} G(w)\Big|_{w=-i\ln z}~,}
where the factor of $z^{-h_l}$ takes into account the transformation of the operator $O_l$.  In terms of the functions $f$ and $\ft$,
\eqn\bh{\eqalign{f(1-z)& =\eps_l \ln z +S(w)\big|_{w=-i\ln z}\cr
  \ft(1-z)& =\eps_l \ln z +(\eps_p-\eps_l-\eps'_l)\ln(1-z)+S(w)\big|_{w=-i\ln z} }}

Each geodesic segment is obtained by extremizing the standard worldline action $I = \eps\int\! d\lambda \sqrt{g_{\mu\nu}{dx^\mu \over d\lambda}{dx^\nu \over d\lambda}}$, where we choose $\lambda$ to be proper length. Geodesics thus obey
\eqn\bi{  {1\over \cos^2 \rho} \dot{\rho}^2 +  { p_\phi^2\over \alpha^2}\cot^2 \rho =1 }
where the conserved momentum conjugate to $\phi$ is
\eqn\bj{ p_\phi =\alpha^2 \tan^2 \rho~ \dot{\phi}~.  }
This gives
\eqn\bk{ \cos \rho = {1\over \sqrt{1+p_\phi^2/\alpha^2}}{1\over \cosh \lambda }}
The regulated length is defined by imposing a cutoff near the boundary, at $\cos \rho = \Lambda^{-1}$.

As a simple example, consider the case $\eps'_l = \eps_l$ with $\eps_p \ll \eps_l$.  In this regime we can first work out the geodesic connecting the two boundary points.  Attached to this will be a geodesic connecting the midpoint with $\rho=0$.   To first order in $\eps_p/\eps_l$ we can neglect the fact that the latter geodesic will ``pull" on the former.     A simple computation yields for the $\eps_l$ geodesic
\eqn\bl{ \cos \rho  = {\sin {\alpha w \over 2} \over \cosh \lambda}~.}
Its regulated length is given by
\eqn\bm{ 2L_l = 2\lambda\big|_{\cos \rho = \Lambda^{-1} } = 2 \ln \left( \sin {\alpha w \over 2} \right)+2\ln \left(\Lambda \over 2\right)~.}
The length of the $\eps_p$ geodesic is
\eqn\bn{L_p = \int_0^{\cos \rho = \sin {\alpha w \over 2} } \! {d\rho \over \cos \rho} =- \ln \left(\tan {\alpha w \over 4 } \right)~.    }
The contribution to the correlator is then
\eqn\bo{ G(w) = e^{-2h_l L_1 - h_p L_p} =  { \left( \tan {\alpha w \over 4}\right)^{h_p} \over \left( \sin {\alpha w \over 2}\right)^{2h_1} }~,  }
where we dropped an uninteresting $w$-independent prefactor.\foot{This includes the dependence on the regulator $\Lambda$.  Here and elsewhere we simply drop such regulator dependent terms, since they contribute no w-dependence.}  After a little bit of algebra, we then find
\eqn\bp{ f(1-x)= \left[\ln x +2 \ln \left( x^{-{\alpha\over 2}}- x^{{\alpha\over 2}}\over \alpha\right)  \right] \eps_l -  \ln \left( {x^{-{\alpha\over 4}}-x^{{\alpha\over 4}}\over 4\alpha( x^{-{\alpha\over 4}}+x^{{\alpha\over 4}})} \right)\eps_p~,  }
in agreement with \az.  Note that although we assumed $\eps_p \ll \eps_l$,  since the result is linear it turns out to agree with \az, where no such assumption was made.
 
Although we mainly focus on $\alpha>0$, the result \bp\ makes sense in the $\alpha \rt 0$ limit.  In this case the bulk metric becomes Poincar\'{e} AdS after a coordinate rescaling.   The $\eps_p$ geodesic now disappears down the infinite throat towards the Poincar\'{e} horizon. 

In the above, the worldines were taken to extremize the action, which implies that the stress tensor of the particles is covariantly conserved. This is needed in order that the particles can consistently couple to gravity, a fact that we will need later when we explain the relation between the geodesic approach and the monodromy approach.   In this regard, we also note that we only need to require that the worldlines extremize the action, and they need not furnish a global minimum.

\subsec{Result for general light operators}

\def\ep{\epsilon}
\def\frac#1#2{{#1\over #2}}
\def\dd#1#2{\frac{d #1}{d #2}}
\def\pp#1#2{\frac{\partial #1}{\partial #2}}
\def\lr#1{\left( #1 \right)}
\def\inv#1{\frac{1}{#1}}

Working in a fixed-$t$ slice, our goal is to find the minimum value of the worldline action \bez{} as a function of the angular separation $\Delta\phi$ of the light operators $O_l$, $O_{l'}$. We are interested in the regime where the action is minimized by a configuration in which all three worldlines' lengths are nonzero. This requires each of the three weights $\ep_l,\ep_l',\ep_p$ to be less than the sum of the other two. In the complementary case, where one of the three weights is greater than or equal to the sum of the other two, it is simple to minimize the action but the relation to the CFT picture breaks down.

By varying the action with respect to the location of the cubic vertex, one finds that at the vertex
\eqn\ftot{
\ep_l\dot{x}^{\mu}_l +\ep_l'\dot{x}^{\mu}_{l'} + \ep_p\dot{x}^{\mu}_p=0
}
where the dots denote derivatives with respect to a proper length parameter that increases away from the vertex. The problem of computing the minimum value of $S$ is identical to that of computing the energy in equilibrium of three elastic ``rubber bands'' whose tensions are $\ep_l,\ep_l',\ep_p$, independent of their lengths. Equation \ftot{} is the equilibrium condition of vanishing net force on the cubic vertex.

The angular and radial components of equation \ftot{} are
\eqna\fbal
$$\eqalignno{
\ep_lp_{\phi}+\ep_l'p_{\phi}' &=0 &\fbal a\cr
\ep_l\sqrt{1-\frac{p_{\phi}^2}{\alpha^2\tan^2\rho}}
+
\ep_l'\sqrt{1-\frac{{p_{\phi}'}^2}{\alpha^2\tan^2\rho}}
&=\ep_p &\fbal b\cr
}$$
where $\rho$ is the radial coordinate of the cubic vertex and
\eqn\defP{
p_{\phi} = \alpha^2(\tan^2\rho_l)\dot{\phi_l}
}
is the conserved momentum along worldline $l$ that comes from the metric's $\phi$-translation isometry, with $p_{\phi}'$ defined similarly.

The sign of each square root in equation \fbal{b} is positive if the corresponding worldline approaches the cubic vertex from radially outward and negative if it approaches from inward. When $|\ep_l^2-\ep_l'^2|<\ep_p^2$ equations \fbal{a,b} can only be true if both square roots are positive, meaning both worldlines must approach the cubic vertex from the outward direction. On the other hand when $|\ep_l^2-\ep_l'^2|>\ep_p^2$ equations \fbal{a,b} require the worldline corresponding to the smaller of $\ep_l,\ep_l'$ to approach the vertex from the inward direction.

Equations \fbal{a,b} also imply
\eqn\cstreq{
\frac{(p_{\phi}\ep_l)^2}{\alpha^2\tan^2\rho}
=\frac{(p_{\phi}'\ep_l')^2}{\alpha^2\tan^2\rho}
= \mu^2
}
where $\mu>0$ is defined by
\eqn\defmu{
\mu^2=\frac{\ep_l^2+{\ep_l'}^2-\ep_p^2/2}{2} - \frac{(\ep_l^2-{\ep_l'}^2)^2}{4\ep_p^2}.
}
$\mu^2$ is positive as a consequence of the assumption that each of the three weights $\ep_l,\ep_l',\ep_p$ is less than the sum of the other two.

In light of equation \fbal{a} it is useful to characterize the shapes of geodesics $l$, $l'$ with a single parameter rather than the redundant set $(p_{\phi},p_{\phi}')$. We define
\eqn\defP{
P := \frac{\ep_l}{\alpha\mu}p_{\phi} = -\frac{\ep_l'}{\alpha\mu}p_{\phi}'
}
and assume without loss of generality that $P\geq 0$. Equation \cstreq{} implies that $P$ is in fact the tangent of the $\rho$ coordinate of the cubic vertex.

When both worldlines approach the vertex from outward, the angular separation $\Delta\phi$ between their endpoints is
\eqn\deltaphiofP{
\Delta\phi = \frac{\mu}{\alpha \ep_l}
\int_{1}^{\infty}\frac{du}{u\sqrt{1+(Pu)^2}\sqrt{u^2-(\mu/\ep_l)^2}} + (\ep_l\to\ep_l')~.
}
The integration variable $u$ is related to the radial coordinate $\rho$ by $u=(\tan\rho)/P$. When one of the worldlines approaches the vertex from inward instead, one must add to the right hand side of equation \deltaphiofP{} the angle that worldline sweeps out while its radial coordinate is smaller than that of the cubic vertex. An expression for that angle is
\eqn\extrapiece{
\phi_{(\rho\,<\,\arctan P)}=\frac{2\mu}{\alpha \ep_s}
\int_{\mu/\ep_{s}}^{1}\frac{du}{u\sqrt{1+(Pu)^2}\sqrt{u^2-(\mu/\ep_s)^2}}~,
}
where $\ep_s$ is the smaller of $\ep_l,\ep_l'$.

We will need to solve equation \deltaphiofP{} for $P$ as a function of $\Delta\phi$. One can put the equation in the form
\eqn\expalphaphi{
e^{i\alpha\Delta\phi} =
\lr{\frac{\cos\gamma\cos\psi+i\sin\psi}{\cos\gamma+i\sin\gamma\sin\psi}}
\lr{\frac{\cos\gamma'\cos\psi+i\sin\psi}{\cos\gamma'+i\sin\gamma'\sin\psi}}~,
}
where the angles $\psi,\gamma,\gamma'$ are all between $0$ and $\pi/2$ and are defined by $\cot\psi = P$, $\cos\gamma=\mu/\ep_l$, $\cos\gamma'=\mu/\ep_l'$.

We may trade out $\psi$ for a new variable $z$ defined by $\cos\psi=(z+z^{-1})/2$, $\sin\psi=(z-z^{-1})/2i$. Equation \expalphaphi{} is equivalent to
\eqn\quadr{
e^{i\alpha\Delta\phi} = \lr{\frac{z+z\cos\gamma+\sin\gamma}{1+\cos\gamma+z\sin\gamma}}
\lr{\frac{z+z\cos\gamma'+\sin\gamma'}{1+\cos\gamma'+z\sin\gamma'}}~.
}
This is a  quadratic equation for $z$.  It can be rewritten as a quadratic equation for $P$ and solved to yield
\eqn\Poftheta{
P = \frac{\ep_l+\ep_l'}{2\mu}\cot\theta
-\frac{\sqrt{\ep_p^2-(\ep_l-\ep_l')^2\sin^2\theta}}{2\mu\sin\theta}~.
}
We have introduced $\theta := \alpha\Delta\phi/2$.

The expression for $P$ in equation \Poftheta{} gives, via equation \defP{}, the conserved momenta $p_{\phi}$, $p_{\phi}'$ of worldlines $l$, $l'$ in terms of the angular separation $\Delta\phi$ between their endpoints. In the case where one worldline approaches the cubic vertex from inward equation \Poftheta{} continues to hold.

Now, the variation of the action \bez\ with respect to the locations $x^{\mu}_l$, $x^{\mu}_{l'}$ of the worldlines' boundary endpoints is
\eqn\deltaS{
d S = \ep_l g_{\mu\nu}\dot{x}^{\mu}_l d x^{\nu}_l
+\ep_{l}' g_{\mu\nu}\dot{x}^{\mu}_{l'} d x^{\nu}_{l'}.
}
Here the dot denotes a derivative with respect to a proper length parameter that decreases away from the boundary. (In the picture where the worldlines are rubber bands and $S$ is their energy, the right hand side of equation \deltaS{} is the work required to move the ends of the rubber bands.) It follows that the change in the action from a small increase in the angular separation $\Delta\phi$ of the wordlines' endpoints is
\eqn\deltaStwo{
d S = \alpha \mu P d\Delta\phi~.
}
The endpoints are at the locations of the light operators, \bda. Their angular separation is $w$, and so
\eqn\deltaSthree{
\pp{S}{w} = \alpha \mu P
}
where $P$ is given by equation \Poftheta{}, and the variable $\theta$ is related to $w$ by $\theta=\alpha w/2$. Equation \deltaSthree{} can be integrated to give
\eqn\Sofw{
S(w) = (\ep_l+\ep_l')\ln\sin\theta + \ep_p\rm{arctanh}\frac{\cos\theta}{\sqrt{1-\beta^2\sin^2\theta}}-|\beta|\ep_p\ln\lr{|\beta|\cos\theta + \sqrt{1-\beta^2\sin^2\theta}}
}
where $\beta := (\ep_l'-\ep_l)/\ep_p$.

Plugging this into \bh, it yields
\eqn\fgenlight{\eqalign{
f(1-z) &= \ep_l\ln z + (\ep_l+\ep_l')\ln\sin\theta + \ep_p\rm{arctanh}\frac{\cos\theta}{\sqrt{1-\beta^2\sin^2\theta}}\cr
& \quad -|\beta|\ep_p\ln\lr{|\beta|\cos\theta + \sqrt{1-\beta^2\sin^2\theta}}
}}
with
\eqn\sinesub{ \cos \theta = { z^{\alpha/2}+z^{-\alpha/2} \over 2}~,\quad \sin \theta = { z^{\alpha/2}-z^{-\alpha/2} \over 2i}~. }
This result can be verified by checking that it agrees with the result from the monodromy approach, and also by expanding in $z$ and verifying agreement with the result of the recursion relation.  Also, it is straightforward to verify that upon setting $\ep_l'=\ep_l$ we recover \az.

The results in this section hold for the special case that the endpoints of the $\ep_l$ and $\ep'_l$ geodesics lie on a common time slice on the boundary.  In appendix B we study the generalization to unequal times and explain how the correct conformal blocks emerge in this case as well.

\newsec{Relation between geodesic and monodromy approaches}

In this section we explain why computing the action for bulk geodesics gives answers for the semi-classical conformal blocks that agree with those of the monodromy approach.    The argument is very simple.  Given a geodesic configuration, we can work out the linearized metric perturbation sourced by the particles. This perturbation gives rise to a boundary stress tensor that can be identified with $T(z)$ appearing in the monodromy method.   The monodromy conditions arise by requiring that Einstein's equations are obeyed at the geodesics.  Each geodesic segment carries a conserved momentum $p_\phi$, which can be identified with the accessory parameter $c_x$, since it appears as the residue of a simple pole in $T(z)$.  Finally, $p_\phi = {d S \over d\phi}$, just as $c_x$ is related to the derivative of $f$.  This establishes the relation between the geodesic action $S$ and the function $f$ appearing in the monodromy method.

We first show how to relate the simple pole in $T(z)$ to $p_\phi$.   We consider a metric with the usual Fefferman-Graham expansion near the boundary,
\eqn\daa{ds^2 = d\rho^2 + e^{2\rho}g^{(0)}_{\mu\nu} dx^\mu dx^\nu + g^{(2)}_{\mu\nu}dx^\mu dx^\nu + \ldots }
and take $g^{(0)}_{\mu\nu} dx^\mu dx^\nu  = dw d\wb$.  We also include a particle with worldline action
\eqn\db{ S_h = 2h \int\! d\lambda \sqrt{g_{\mu\nu}{dx^\mu \over d\lambda}{dx^\nu \over d\lambda}}~,}
with $\lambda$ equal to proper length.    We consider the case that the worldline pierces the boundary at some location $w_0$, and we wish to consider the Einstein equations near this point.  Expanding the Einstein equations for large $\rho$ to first order, the only non vanishing equations read
\eqn\dc{\eqalign{ g^{(2)}_{w\wb} & =2\pi   h \delta^{(2)}(w-w_0)  \cr
 \p_{\wb} g^{(2)}_{ww} - \p_w g^{(2)}_{w\wb} & =-4\pi {p_w}  \delta^{(2)}(w-w_0) \cr
 \p_{w} g^{(2)}_{\wb\wb} - \p_{\wb} g^{(2)}_{w\wb} & =-4\pi  {p_{\wb}}  \delta^{(2)}(w-w_0) }}
where the (rescaled) mass and   canonical momentum are ${c\over 6}\eps=h$ and ${c\over 6} p^\mu = 2h {dx^\mu \over d\lambda}$, $c$ being the usual Brown-Henneaux central charge, $c=3\ell/2G$.  Now, the components of $g^{(2)}_{\mu\nu}$ are just the boundary stress tensor,  $g^{(2)}_{\mu\nu}=T_{\mu\nu}$  (rescaled as in \ad).  Thus,
\eqn\dd{\eqalign{\p_{\wb} T & = -2\pi \eps \p_w \delta^{(2)}(w-w_0) +4\pi p_w\delta^{(2)}(w-w_0) \cr
\p_{w} \Tb & = -2\pi \eps \p_{\wb} \delta^{(2)}(w-w_0) +4\pi  p_{\wb} \delta^{(2)}(w-w_0)~.}}
Using $\p_{\wb} {1\over w} = 2\pi \delta^{(2)}(w)$, and $p_w = {p_\phi\over 2}$ we find
\eqn\de{\eqalign{ T(w) & = {\eps\over (w-w_0)^2} + {p_\phi \over w-w_0} +\ldots \cr
\Tb(\wb) & = {\eps\over (\wb-\wb_0)^2} + {p_\phi \over \wb-\wb_0} +\ldots }}
where $\ldots$ denote non-singular terms.  By the usual relation between canonical momentum and the variation of the action under a change of boundary conditions we have ${c\over 6} p_\phi = {dS_h \over d\phi}$.   The relation to the formulas appearing in monodromy approach is now clear: just as $c_x$ appeared as the residue of the simple pole in the stress tensor and was related to derivative of $f$, the same is true of $p_\phi$, now related to the derivative of the geodesic action.

The final step to demonstrate the equivalence of the two approaches is to to show how the monodromy conditions arise in the bulk.  We first note that the full bulk metric will take the form
\eqn\df{ ds^2 = d\rho^2 - Tdw^2 - \Tb d\wb^2 +(e^{2\rho}+T \Tb e^{-2\rho})dwd\wb~.}
This is a solution of the source free Einstein's equations if $\p_{\wb}T= \p_w \Tb=0$.  At the location of the particle worldlines these equations are corrected, as in \dd.

We now pass to the Chern-Simons formulation of 2+1 gravity with negative cosmological constant.  This was employed in a closely related context in \deBoerSNA.  The metric is replaced by an SL(2) $\times$ SL(2) connection,
\eqn\dg{\eqalign{ A = \left(\matrix{{1\over 2} d\rho & e^{-\rho}T dw \cr
-e^\rho dw & -{1\over 2}d\rho   }\right)~,\quad \Ab = \left(\matrix{-{1\over 2} d\rho & e^{\rho} dw \cr
-\Tb e^{-\rho} dw & {1\over 2}d\rho   }\right)~. }}
Now consider the holonomy of this connection around a closed contour $C$, which we take to lie at fixed $\rho$. Focussing just on $A$,
\eqn\dh{ {\rm Hol}[C] = P e^{\oint A}~.}
In the absence of matter the holonomy would be trivial, since Einstein's equations are equivalent to flatness of the connections.   But for a contour that encircles a particle worldline, the holonomy will pick up a contribution fixed by the mass of the particle.    To relate this to the monodromy approach, we note that computing ${\rm Hol}[C]$ is equivalent to computing the monodromy of the system of differential equations
\eqn\di{   {d\psi \over dw} = A \psi}
where $\psi$ is a two component vector.  The bottom component obeys
\eqn\dj{ \psi_2'' +T \psi_2 =0~,}
which we recognize as the ODE appearing in \an.

We focus on a contour that encircles the two operator insertion points at $0$ and $w$.  This contour encircles the worldline corresponding to the exchanged primary $O_p$.   It is then clear that if $T$ is such that the monodromy of the differential equation \dj\ is related to $\eps_p$ in the correct way, then the holonomy of the Chern-Simons connection will be such that we solve the Einstein equations in the presence of the particle worldline.

Summarizing, we see that given a linearized solution of Einstein's equations in the presence of particle worldlines we can find a solution of the monodromy problem.  Further, the action of the bulk solution agrees with the function $f$ appearing in the exponent of the semi-classical conformal block.

\newsec{Discussion}

We have achieved a clean AdS$_3$ bulk interpretation of the semi-classical conformal blocks, extending the observations in \FitzpatrickVUA.  To linear order in the light operator dimensions, we simply have to find the equilibrium configuration of three geodesic segments joined at cubic vertex, propagating in a geometry dual to the heavy operators.       We close with a couple of observations and questions for the future.

It is straightforward in principle to go to higher orders in the light operator dimensions.   From the bulk point of view, this just corresponds to solving Einstein's equations order by order in Newton's constant.  Phrased in the language of the Chern-Simons formulation, the problem consists of finding locally flat connections with specified holonomies around contours representing the locations of the worldlines.   As at linear order, this is the same problem as in the monodromy approach.

It should also be straightforward to consider conformal blocks for higher point functions. In this case we would have more worldlines attached with additional cubic vertices.  Conformal blocks on higher genus Riemann surfaces could also be considered.

We have focused here on the semi-classical conformal block, which is the leading term in the large $c$ expansion.  More generally, we can think of writing
\eqn\za{ \Fct(x) = e^{ -{c\over 6} ( \ft^{(0)}(x) + {1\over c} \ft^{(1)}(x) + \ldots)}~.}
In terms of the loop expansion in the bulk, we expect that $\ft^{(1)}$ is given by the effects of 1-loop fluctuations around the classical background.   We can use the recursion relation to compute terms in $\ft^{(1)}(x)$.   For example, setting $\eps'_l=\eps_l$ and $\eps_p=0$, we find, at linear order in $\delta$,
\eqn\zb{\eqalign{ \ft^{(1)}(x) &={3\over 16}(4\eps_h-1) x^2 +{3\over 16}(4\eps_h-1) x^3+{1\over 512}(4\eps_h-1)(4\eps_h+87)x^4 \cr
  & -{3\over 4}(4\eps_h-1)\eps_l x^2 -{3\over 4}(4\eps_h-1)\eps_l x^3  -{1\over 14400}(5296 \eps_h^2+37392 \eps_h -9675)\eps_l x^4 +\ldots  }}
where the $\ldots$ denote terms higher order in $\eps_l$  and in $x$.
It is important to note that to obtain this we first extract the large $c$ asymptotics for general values of $\eps_l$, $\eps'_l$ and $\eps_p$, and only afterwards set $\eps'_l=\eps_l$ and $\eps_p=0$.  The terms in the first line are puzzling, as they are nonzero even upon setting $\eps_l=0$.  It will be interesting to understand their physical origin.

Once the bulk interpretation of the full conformal block is established, including the subleading $1/c$ effects, we can think of using this information to interpret specific CFT correlators.  It will be interesting to apply this to the black hole context, where the problem of information loss can be phrased in terms of such correlators.

Another interesting direction to consider is the extension to higher spin theories, where the Virasoro algebra is enhanced to a $\Wc$-algebra.  In \deBoerSNA\ it was established that the semi-classical vacuum block admits a bulk realization in terms of a Wilson line \refs{\deBoerVCA,\AmmonHBA,\CastroMZA} embedded in an asymptotically AdS$_3$ background with higher spin fields excited.   A natural question is how to extend this story beyond the vacuum block.

\vskip .3in

\noindent
{ \bf Acknowledgments}

\vskip .3cm

We wish to thank Eric Perlmutter for useful conversations.
P.K. is supported in part by NSF grant PHY-1313986.

\appendix{A}{Recursion relation}

The series expansion of the conformal block can be computed using the recursion relation presented in \zamo.    We first write
\eqn\za{ \Fct^{34}_{21}(x) =\left({16 q\over x}\right)^{h_p+{1-c\over 24}} (1-x)^{{c-1\over 24}-h_2 -h_3}\theta_3(q)^{{c-1\over 2}- 4\sum_i h_i}H(c,h_p,h_i,q)  }
where $H(c,h_p,h_i,q)$ is the quantity that will be computed by the recursion relation.    $q$ is related to $x$ by
\eqn\ze{ q=e^{i\pi \tau}~,\quad \tau = i {K(1-x)\over K(x)}~,\quad K(x)= {1\over 2} \int_0^1 {dt \over [t(1-t)(1-xt)]^{1\over 2}}}
or equivalently
\eqn\zf{x = \left({\theta_2(q) \over \theta_3(q)}\right)^4 }
This gives
\eqn\zfa{16 q =x+{1\over 2}x^2 +{21\over 64}x^3 +{31\over 128 }x^4 +{6257\over 32768}x^5+ \ldots }
External conformal dimensions $h_i$ are written in terms of $\lambda_i$ as
\eqn\zg{ h_i = {c-1\over 24} + \lambda_i^2 }
We further define
\eqn\zd{\eqalign{ \alpha_{\pm} &= \sqrt{ 1-c\over 24} \pm \sqrt{25-c\over 24} \cr
 \lambda_{pq}& = \alpha_+ p + \alpha_- q \cr
 \Delta_{mn}(c)& = {c-1\over 24} +{(\alpha_+m + \alpha_- n)^2 \over 4}  }}
The recursion relation is then
\eqn\zb{ H(c,h_p,h_i,q) = 1+ \sum_{m>0,n>0}{(16q)^{mn}R_{mn}(c,h_i)H(c,\Delta_{mn}+mn,h_i,q)\over h_p-\Delta_{mn}(c)} }
with
\eqn\zc{ R_{mn}(c,h_i)=-{1\over 2}{ \prod_{p,q}(\lambda_2+\lambda_1- {\lambda_{pq}\over 2})(\lambda_2-\lambda_1-{\lambda_{pq}\over 2}) (\lambda_3+\lambda_4-{\lambda_{pq}\over 2})(\lambda_3-\lambda_4-{\lambda_{pq}\over 2})  \over \prod'_{k,l} \lambda_{kl}   }      }
The product in the numerator is taken over $p=-m+1, -m+3, \ldots , m-3, m-1$;  $q=-n+1, -n+3, \ldots , n-1$.     The product in the denominator is taken over $k=-m+1, -m+2, \ldots, m$; $l=-n+1, -n+2, \ldots n$, and the prime means that we omit $(k,l)=(0,0)$ and $(k,l)=(m,n)$.

\appendix{B}{Extension to operators at different times}
Equation \bh\ gives the relation between the value of the worldline action \bez\ and the corresponding conformal block when the two light operators lie on the same time slice. In this appendix we argue that the extension of \bh\ to operators at different times is
\eqn\xba{ f(1-z)+\bar{f}(1-\bar{z}) - \eps_l\ln z\bar{z} = 2S(\phi ,\tau) }
where $S(\phi,\tau)$ is the worldline action as a function of the location $(\phi,\tau)$ on the cylinder of the light operator $O_l$, with $O_l'$ fixed at $(\phi,\tau)=(0,0)$. The relation between $(\phi,\tau)$ and $(z,\bar{z})$ is $z=e^{iw},\bar{z}=e^{i\bar{w}}$ with $w=\phi+i\tau$, $\bar{w}=\phi-i\tau$. Recall that $f$ is defined in terms of the holomorphic conformal block $\Fc$ by $\Fc=e^{-{c\over 6}f}$. Similarly $\bar{f}$ is defined in terms of the antiholomorphic conformal block $\bar{\Fc}$ by $\bar{\Fc}=e^{-{c\over 6}\bar{f}}$.

To regularize the worldline action we place the boundary at $\cos\rho=\ep$, where the cutoff $\ep$ is independent of $\phi$ and $\tau$.

Given that equation \bh\ holds for real $w$, the two sides of equation \xba\ agree when $\tau=0$. Their derivatives with respect to $\tau$ agree as well; they both vanish by $\tau\to -\tau$ symmetry. The left hand side of equation \xba\ is the real part of a holomorphic function of $\phi+i\tau$ and therefore satisfies Laplace's equation. Thus if $S(\phi,\tau)$ satisfies Laplace's equation
\eqn\xbb{ \left({\p^2 \over \p \phi^2} + {\p^2 \over \p \tau^2} \right) S(\phi,\tau) = 0}
then equation \xba\ must hold for all values of $(\phi,\tau)$ for which $S(\phi,\tau)$ is defined.

We now want to show that $S(\phi,\tau)$ indeed satisfies \xbb{}. Let $\tilde{S}(x,y)$ be the worldline action as a function of the location $x$ of operator $O_l$ and the location $y$ in AdS$_3$ of the cubic vertex. Let $x_0$ be given and let $y_0$ be the $y$ that minimizes $\tilde{S}(x_0,y)$.  Under the displacement $(x_0,y_0)\rt (x_0+dx,y_0+dy )$  the action $\tilde{S}$ becomes, to quadratic order in the displacements,
\eqn\xbc{\eqalign{
\tilde{S}(x_0+dx,y_0+dy)=&S(x_0)-\ep_lv_{\mu}dx^{\mu} + \inv{2}K_{\mu\nu}dy^{\mu}dy^{\nu}\cr
&+ \ep_l\lr{dx^{\mu}dx^{\nu}\lr{e^{-2L_l}g_{\mu\nu} + \frac{1}2 v_{\mu}v_{\nu}}-2e^{-L_l}\lr{g_{\mu\nu}-v_{\mu}v_{\nu}}dx^{\mu}dy^{\nu}}. }}
Every tensor on the right hand side lives at the point $x_0$. In particular, $dy^{\mu}$ is the parallel transport of $dy$ to the point $x_0$ and $g_{\mu\nu}$ is the metric at $x_0$. $v_{\mu}$ is the unit vector pointing from $x_0$ down the geodesic toward $y_0$, and $L_l$ is the length of that geodesic. Because $x_0$ is on the boundary $e^{-L_l}$ is proportional to the cutoff, $\ep$. Equation \xbc\ is true to zeroth order in the cutoff.

The term $\inv{2}K_{\mu\nu}dy^{\mu}dy^{\nu}$ captures the change in the worldline action from the changes in lengths of geodesics $l'$ and $p$ and also the part of the length change of geodesic $l$ that is independent of $dx$. The explicit form of $K$ is
\eqn\xbd{
K_{\mu\nu} = \eps_l(2g_{\mu\nu}-3v_{\mu}v_{\nu})
+\eps'_l(2g_{\mu\nu}-3v'_{\mu}v'_{\nu})
+\eps_p\sin\rho_0 v^p_{\mu}v^p_{\nu} }
where the unit vectors $v',v^p$ point from the cubic vertex down the corresponding geodesics and have been parallel transported to $x_0$, and $\rho_0$ is the $\rho$ coordinate of the vertex.

Given a particular $dx$, the function $\tilde{S}(x_0+dx,y_0+dy)$ is minimized for some particular value of $dy$, call it $dy_*$, which is the solution to the linear equation
\eqn\xbe{K_{\mu\nu}dy_*^{\nu}=2\ep_l e^{-L_l}\lr{g_{\mu\nu}-v_{\mu}v_{\nu}}dx^{\nu}.}
Substituting for $dy_*$ in $\tilde{S}(x_0+dx,y_0+dy_*)$ gives the minimized worldline action at $x=x_0+dx$ to second order in $dx$:
\eqn\xbf{\eqalign{S(x_0+dx) = &S(x_0) -\ep_lv_{\mu}dx^{\mu} + \ep_l dx^{\mu}dx^{\nu}\lr{e^{-2L_l}g_{\mu\nu} + \frac{1}2 v_{\mu}v_{\nu}}
\cr 
&- \ep_l e^{-2L_l}dx^{\mu}\lr{g_{\mu\rho}-v_{\mu}v_{\rho}}(K^{-1})^{\rho\sigma}(g_{\sigma\nu}-v_{\sigma}v_{\nu})dx^{\nu}}}
where $(K^{-1})^{\mu\sigma}K_{\sigma\nu} = \delta^{\mu}_{\nu}$. From equation \xbf\ one can read off
\eqn\xbg{\nabla_{\mu}\nabla_{\nu}S(x) = \ep_l \lr{2e^{-2L_l}g_{\mu\nu}+v_{\mu}v_{\nu}}- 2\ep_l e^{-2L_l}\lr{g_{\mu\rho}-v_{\mu}v_{\rho}}(K^{-1})^{\rho\sigma}(g_{\sigma\nu}-v_{\sigma}v_{\nu}).}
The point $x_0$ is on the boundary cylinder, and so $g_{\mu\nu}$, $v_{\mu}v_{\nu}$, $K_{\mu\nu}$ are all of order $\ep^{-2}$, in the sense that their components in the $(\rho,\phi,\tau)$ coordinate system are of order $\ep^{-2}$. $g^{\mu\nu}$ and $(K^{-1})^{\mu\nu}$ are both of order $\ep^2$. Keeping only the most divergent terms in equation \xbg{} we find
\eqn\xbh{\nabla_{\mu}\nabla_{\nu}S(x) = \eps_l v_{\mu}v_{\nu} + O(\ep^{-1} ).}
Finally, we restrict to displacements $dx$ that keep $x$ on the boundary cylinder. Letting $n^{\mu}$ be the unit inward-pointing normal vector at $x$, the two-dimensional Laplacian of $x$ is
\eqn\xbi{\nabla^2 S = \lr{g^{\mu\nu}-n^{\mu}n^{\nu}}\nabla_{\mu}\nabla_{\nu}S(x)=\ep_l\lr{1-(n\cdot v)^2} + O(\ep ).}
To lowest order in $\ep$ the quantity $n\cdot v$ is unity, and so
\eqn\xbj{\nabla^2 S = 0+ O(\ep ).}
Thus the regularized worldline action satisfies Laplace's equation, which concludes the proof of equation \xba{}.

\listrefs
\end